# Electrically stabilized magnetic vortex and antivortex states in magnetic dielectrics

A.P. Pyatakov[*], G.A. Meshkov

*Physics Department, M.V. Lomonosov MSU, Leninskie gori, Moscow, 119991, Russia*

The micromagnetic distribution in a dielectric nanoparticle is theoretically considered. It is shown that the existence of inhomogeneous magnetoelectric interaction in magnetic dielectrics provides the possibility to stabilize the vortex and antivortex state in magnetic nanoparticle. The estimation of the critical voltage necessary for vortex/antivortex nucleation in bismuth ferrite and iron garnet nanoparticles yields a value of ±150 V. This system can be considered as electrically switchable two state-logic magnetic element.

*) pyatakov@physics.msu.ru

Numerous micromagnetic structures observed in magnetic media are the result of competition between only a few types of interactions that include magnetostatic and exchange energy. These two provide us with magnetic vortex structures in ferromagnetic nanodiscs or nanodots [1-6]. An equally fundamental though much less known is antivortex state that is a topological counterpart of the vortex. The realization of antivortex state is a challenging task since it is unfavorable for magnetostatic reasons due to the formation of the magnetic charges at the edge of the particle. There are only a few reports in which the antivortex was observed either in a complicated four ferromagnetic rings cross-junction structure [7] or as a metastable state of asymmetric cross-like nanostructures [8].

Meanwhile there is an additional interaction that should be taken into account when we consider micromagnetic structure under influence of electric field or the structures in magnetic ferroelectrics (multiferroics). It is proportional to spatial derivatives of magnetic order parameter vector. This *spin flexoelectricity* is the variety of magnetoelectric coupling related to the magnetic inhomogeneties and it is described by $P_i M_j \nabla_k M_n$ coupling term, where P and M, are electric and magnetic order parameters, respectively [9-18]. This type of interaction is responsible for magnetically induced electric polarization in spiral multiferroics [19] and electric field driven magnetic domain wall motion in ferrimagnetic iron garnet films [20].

In this short Letter we will show that the gradient electric field produced by point electrode (e.g. cantilever tip of atomic force microscope) can stabilize in magnetic dielectric nanoparticle either vortex or antivortex state depending on the electric polarity of the tip.

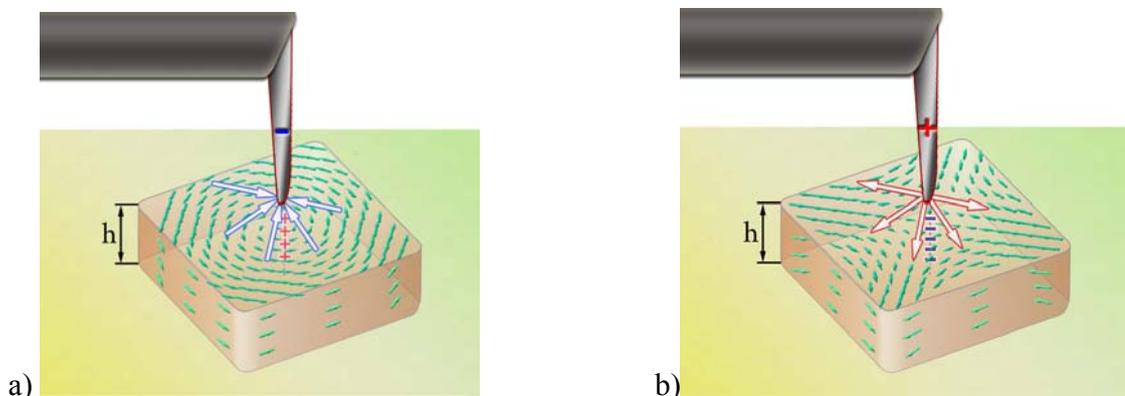

a)          b)

Fig. 1 Magnetic dielectric nanoparticle subjected to the electric field from the cantilever tip (the electric polarity depends on the sign of magnetoelectric constant in (1))
a) electrically induced magnetic vortex state b) electrically induced magnetic antivortex state.

Magnetoelectric materials, mostly being ferrimagnets or weak ferromagnets, have rather moderate value of spontaneous magnetization. This results in large exchange length $l_{exch} = \sqrt{2A/M_s^2}$, where $A$ is the exchange stiffness, $M_S$ is saturation magnetization, and large magnetostatic lengths $l_{MS} = 2\sqrt{AK}/M_s^2$ (the maximum size of single domain particle, where $K$ is anisotropy constant) compared to conventional ferromagnets. The typical values for room temperature magnetoelectric iron garnet films are: $l_{exch} = 0,5 \mu m$, $l_{MS} = 5 \mu m$ and room temperature multiferroic bismuth ferrite: $l_{exch} = 1.5 \mu m$, $l_{MS} = 240 \mu m$. Thus the magnetic dielectric particles of nanometric size should be in homogeneous magnetic state since exchange interaction is the predominant one.

The only thing that can compete against the exchange interaction in magnetic dielectrics on this nanometer scale is the spin flexoelectricity that for isotropic media can be represented in the form [10,13]:

$$F_{ME} = \gamma \cdot \mathbf{P} \cdot (\mathbf{n} \cdot (\nabla \cdot \mathbf{n}) - (\mathbf{n} \cdot \nabla)\mathbf{n}) \tag{1}$$

where $\gamma$ is magnetoelectric constant, $\mathbf{P}$ is electric polarization, $\mathbf{n}$ is unit vector of magnetic order parameter.

The linear charge density corresponding to magnetic vortex can be found as in [13]:

$$Q_L = \pm 2\pi\gamma\chi_e, \tag{2}$$

where $\chi_e$ is electric susceptibility of the medium, signs "±" correspond to the vortex and antivortex state, respectively.

The magnetoelectric energy thus can be represented in the form of electrostatic energy of the charge (2) in the electric potential $\varphi$ of the point electrode:

$$W_{ME} = q\varphi = 2\pi\gamma\chi_e \cdot h \cdot \varphi, \tag{3}$$

where q is the integral charge, $h$ is the height of the particle.

The exchange energy can be estimated as

$$W_{Exch} = Ak^2 V = A\left(\frac{2\pi}{\Delta}\right)^2 h\Delta^2 = A(2\pi)^2 h \tag{4}$$

where $k$ is the modulation wave vector, $\Delta$ is the lateral size of the particle.

Thus the critical potential of the electrode tip in which the nucleation of the vortex (antivortex) occurs that correspond to the condition $W_{ME} + W_{exch} < 0$ can be found as

$$|\varphi_C| = \frac{2\pi A}{\gamma \cdot \chi_e} \tag{5}$$

Assuming $\gamma \sim 10^{-6}$ $\sqrt{erg/cm}$ (typical values for BiFeO$_3$ and iron garnet), dielectric susceptibility $\chi_e \sim 4$ (it can be estimated using the value of BiFeO$_3$ dielectric constant $\varepsilon = 1 + 4\pi\chi_e = 50$) and exchange stiffness $A \sim 3 \ast 10^{-7}$ erg/cm (typical value for room temperature magnets) we obtain $\varphi_c = \pm 150$V.

More rigorous approach should take into account the finite size of the vortex core in which the spin come out of the plane. In the case of dielectric nanoparticle its diameter is determined by the balance between the exchange energy and the inhomogeneous magnetoelectric energy. The estimation (5) remains valid provided that the critical potential corresponds to the lowest voltage at which the vortex/antivortex state appears at the perimeter of the particle, the charge (3) is distributed on the surface of the cylinder wrapping the vortex core and the parameter $\Delta$ in (4) stands for the core diameter.

Thus the electric field can stabilize the vortex state for one sign of the charge and antivortex state for another (see figure). This possibility besides fundamental interest can be considered as a prototypical example of electrically switchable magnetic system with two logic states. The use of ferroelectric media as a material of the tip electrode will make it possible to implement the memory cell based on this principle.

Authors are grateful to A. Zvezdin, D. Khomskii and M. Mostovoy for the interest to the work and valuable discussion.